\documentclass[oneside]{article}

\usepackage[paper=letterpaper,lmargin=2.5cm,tmargin=2.5cm%
,bmargin=2.5cm,rmargin=2.5cm]{geometry}

\usepackage{graphicx}

\usepackage[bottom,flushmargin]{footmisc}

\usepackage{paralist}
\usepackage{amsmath}
\usepackage{amssymb}
\usepackage{array}

\usepackage{amsthm}

\usepackage{float}
\usepackage{flafter}
\usepackage[width=\linewidth]{caption}

\newtheorem*{definition*}{Definition}

\newtheorem{theorem}{Theorem}
\newtheorem{proposition}{Proposition}
\newtheorem*{proposition*}{Proposition}
\newtheorem{problem}{Problem}

\newlength\graphicswidth
\DeclareMathOperator*{\argmin}{argmin}

\graphicspath{{.}{graphics/}}

\begin{document}

\title{Extensions to Network Flow Interdiction on Planar Graphs}

\author{Rico Zenklusen\\
\small Institute for Operations Research, ETH Zurich\\
\small rico.zenklusen@ifor.math.ethz.ch
}
\maketitle


\begin{abstract}

Network flow interdiction analysis studies by how much the value of a 
maximum flow in a network can be diminished by removing components
of the network constrained to some budget. Although this problem is
strongly NP-complete on general networks, pseudo-polynomial algorithms were found
for planar networks with a single source and a single sink and without the
possibility to remove vertices. In this work we introduce
pseudo-polynomial algorithms which overcome some of the restrictions of
previous methods. We propose a planarity-preserving
transformation that allows to incorporate vertex removals and vertex
capacities in pseudo-polynomial
interdiction algorithms for planar graphs. Additionally, 
a pseudo-polynomial algorithm is introduced for the problem of determining
the minimal interdiction budget which is at least needed to make it
impossible to satisfy the demand of all sink nodes, on planar networks
with multiple sources and sinks satisfying that the sum of the supplies at
the source nodes equals the sum of the demands at the sink nodes.
Furthermore we show that the $k$-densest subgraph problem on planar graphs
can be reduced to a network flow interdiction problem on a planar graph
with multiple sources and sinks and polynomially bounded input numbers.
However it is still not known if either
of these problems can be solved in polynomial time.
\end{abstract}

\section{Introduction}

In this paper we are interested in minimizing the maximum flow
of a network by removing arcs and vertices constrained to some
interdiction budget. This problem is known as \emph{network interdiction}
or \emph{network flow interdiction} \footnote{The term \emph{network inhibition}
is also used.}. One can
either allow or disallow partial removal of arcs (removing half of an edge 
corresponds to reduce its capacity to half of the original value).
However the techniques and results seem not to substantially differ
on this issue. We are interested in the case without partial arc destruction.

The problem of finding the \emph{$k$ most vital arcs} of a flow network
is a special case of the network interdiction problem where $k$
arcs have to be removed such that the maximum flow is reduced as
much as possible.
Network interdiction and related problems appear in various areas
such as drug interdiction \cite{wood_1993_deterministic},
military planning \cite{ghare_1971_optimal}, protecting electric
power grids against terrorist attacks \cite{salmeron_2004_analysis}
and hospital infection control \cite{assimakopoulos_1987_network}.

The network interdiction problem was shown to be strongly NP-complete
on general graphs and weakly NP-complete when restricted to planar
graphs \cite{phillips_1993_network,wood_1993_deterministic}.
Different algorithms for finding exact solutions
were proposed (\cite{ghare_1971_optimal,mcmasters_1970_optimal,
ratliff_1975_finding,wood_1993_deterministic})
which are mainly based on branch and bound procedures. In
\cite{burch_2002_decompositionbased} a pseudo-approximation was presented.
Earlier work include \cite{wollmer_1964_removing}.

However when dealing with planar graphs with a single source and sink
it was shown that by using planar duality, pseudo-polynomial
algorithms for the network interdiction problem can be constructed
when only arc removal is allowed \cite{phillips_1993_network}.  
Two of the major drawbacks of these algorithms (apart from the
fact that they can only be applied on planar graphs) are the
restrictions that only arc removals are allowed and that the network
must have exactly one source and one sink.
Vertex removal can easily be formulated as edge removal by a standard
technique of doubling vertices and multiple sources and sink are generally
handled by the introduction of a supersource and supersink
\cite{ahuja_1993_network,ford_1956_maximal}.
However, these transformations destroy
planarity and make it impossible to profit from the currently known
specialized interdiction algorithms for planar graphs.

In this work, we are interested in the development of pseudo-polynomial
algorithms for planar graphs which overcome some of the restrictions
of previous methods. Inspired by algorithms presented in
\cite{khuller_1994_flow,miller_1995_flow}, we propose a planarity-preserving
transformation that allows to incorporate vertex removals and vertex capacities
in pseudo-polynomial interdiction algorithms for planar graphs. Additionally, 
a pseudo-polynomial algorithm is introduced for the problem of determining
the minimal interdiction budget needed to make it
impossible to satisfy the demand of all sink nodes, on planar networks
with multiple sources and sinks satisfying that the sum of the supplies at
the source nodes equals the sum of the demands at the sink nodes.

This problem is closely related to the problem of determining if a flow
network is $n-k$ secure, i.e. any removal of $k$ of its components
does not infect the value of the maximum flow, and can be seen as a
special case of network interdiction.

It is not known if network interdiction on planar networks with 
multiple sources and sinks is a strongly NP-complete problem. We show
that the $k$-densest subgraph problem on planar graphs, a problem for which
neither a polynomial algorithm is known nor is it known if it is NP-complete,
can be reduced to a planar network interdiction problem
with polynomially bounded numbers as input.

The paper is organized as follows. We begin by giving some definitions
and notations in Section~\ref{sec:def_and_not}. In Section~\ref{sec:complexity},
different known complexity results on network interdiction are stated and
we show how the $k$-densest subgraph problem on planar graphs 
can be reduced to a planar network interdiction problem with small input
numbers. Section~\ref{sec:duality_and_algos} presents a simple extension
of currently known algorithms for network interdiction problems on undirected networks were only
arc removals are allowed to the case of directed networks. We present in
Section~\ref{sec:vertex_interdiction} a pseudo-polynomial algorithm
for network interdiction on planar $s$-$t$ flow networks allowing vertex interdiction.
In Section~\ref{sec:pseudopol_nfs_planar} we present a pseudo-polynomial algorithm
for a special network flow interdiction problem with multiple sources and sinks.

\section{Definitions and Notations}\label{sec:def_and_not}

Let $(V,E)$ be a directed graph where $V$ is
the set of vertices, $E$ is the set of arcs and for every arc
$e\in E$, $u(e)\in \{0,1,2,\dots\}$ denotes its capacity.
Two special nodes $s,t\in V, s\neq t$
designate the source node respectively the sink node (the generalization
to multiple sources and sinks is straightforward). We call the network
$G=(V,E,u,s,t)$ a flow network. For $V',V''\subset V$ we denote by
$(V',V'')$ the set of all arcs from $V'$ to $V''$. Furthermore, for
$V'\subset V$ we denote by $\omega^+(V')$ respectively $\omega^-(V')$
the set of all arcs exiting $V'$ respectively entering $V'$, i.e.,
$\omega^+(V')=(V',V\setminus V')$ and $\omega^-(V')=(V\setminus V',V')$
\footnote{We also use the notation $\omega^+_G$ and $\omega^-_G$ to specify that
we are working on the graph $G$.}.
For any subset $V'$ of $V$ we denote by $[V',V\setminus V']$ the
cut defined by $V'$. The value of the cut $[V',V\setminus V']$ 
is $\sum_{e\in \omega^+(V')}u(e)$. In the more general setting when every
arc $e\in E$ has an additional lower bound $l(e)$ on the capacity, the
value of the cut $[V',V\setminus V']$ is defined by
$\nu([V',V\setminus V'])=\sum_{e\in \omega^+(V')}u(e)-\sum_{e\in \omega^-(V')}l(e)$
\footnote{Sometimes the notation $\nu^G$ is used to specify the network.}.
A cut $[V',V\setminus V']$ in $G$ is called \emph{elementary} if the subgraph of $G$ induced by
$V'$ is connected.
For two distinct vertices $s,t\in V$, a cut $[V',V\setminus V']$ is called an $s$-$t$ cut
if $s\in V'$, $t\not\in V'$.

A function $f:E\rightarrow \mathbb{R}$ is called a flow in $G$ (or simply flow
if there is no danger of ambiguity) if it satisfies the following
constraints:
\begin{enumerate}[i)]
\item $0\leq f(e) \leq u(e) \quad\forall\; e\in E$
\item $\sum_{e\in \omega^+(v)}f(e)-\sum_{e\in \omega^-(v)}f(e)=0 \quad
\forall\; v\in V\setminus\{s,t\}$
\item $\sum_{e\in \omega^+(s)}f(e)-\sum_{e\in \omega^-(s)}f(e)\geq0$\;.
\end{enumerate}
For an $s$-$t$
flow $f$ we define its value $\nu(f)$ by
$\sum_{e\in \omega^+(s)}f(e)-\sum_{e\in \omega^-(s)}f(e)$.
A maximum $s$-$t$ flow is an $s$-$t$ flow $f$ with maximum value. The value
of a maximum $s$-$t$ flow in a flow network $G$ is denoted by $\nu^{\max}(G)$.

In the context of network interdiction, for every arc and node
of the network $p\in V\cup E$ an
interdiction cost $c(p)\in \{1,2,3,\dots\} \cup \{\infty\}$ is associated
(with $c(s)=c(t)=\infty$).
The network $G=(V,E,u,s,t,c)$ is called an interdiction network. An interdiction
network has unit interdiction costs if $c(p)\in\{1,\infty\} \; \forall\; p\in V\cup E$.
The network interdiction problem asks to find a set $R\subset V\cup E$
respecting a given budget constraint $c(R):=\sum_{r\in R}c(r)\leq B$ (with
$B\in \{0,1,2,\dots\}$), and among all these sets minimizing the
value of a maximum $s$-$t$ flow on the graph $G\setminus R$, which is the 
subgraph of $G$ obtained by removing the
arcs and vertices contained in $R$ (when removing a vertex,
all arcs adjacent to this vertex are removed too). The value of this
minimum maximum $s$-$t$ flow corresponding to budget $B$ is denoted by $\nu^{\max}_B(G)$
(we therefore have $\nu^{\max}(G)=\nu^{max}_0(G)$).
In this context a set $R\subset V\cup E$ satisfying the budget constraint will be
called an interdiction set. An optimal interdiction set $R$ minimizes the maximum
$s$-$t$ flow with respect to the given budget. Furthermore an optimal interdiction
set $R$ is called minimum if its interdiction cost $c(R)$ is minimum among all
optimal interdiction sets and it is called minimal when removing any arc from
the interdiction set results in a non-optimal interdiction set. 

We define the network flow security problem to be the problem of finding
the minimal budget necessary to decrease the maximum flow by at least one
unit, i.e., $\min\{B\in\{0,1,2,\dots\} \mid \nu^{\max}_B(G)<\nu^{\max}(G)\}$.

The above definitions and problems can easily be extended to interdiction
networks with multiple sources and sinks with fixed supply/demand. 
In this case an interdiction network is given by $G=(V,E,u,S,T,c,d)$ where
$S,T \subset V$ with $S\cap T=\emptyset$ are the set of sources respectively
sinks and the function $d:V\rightarrow \mathbb{Z}$ is the demand/supply
function and satisfies $d(s)<0 \;\forall\: s\in S$, $d(t)>0 \;\forall\: t\in T$ and
$d(v)=0 \;\forall\; v\in V\setminus (S \cup T)$.

To simplify notations a circuit $\mathcal{C}$ in $G$ will be represented by the set of arcs it contains.
For further graph-theoretical terms used in this paper and not further specified in this section we refer to
\cite{korte_2006_combinatorial}.

\section{Complexity}\label{sec:complexity}

\subsection{Previous results}

We associate the following natural decision problems to the network interdiction
problem respectively the network flow security problem.

\begin{problem}[Decision version of network interdiction problem]\label{prob:nip_decision}
Given an interdiction network $G$, some interdiction budget $B\in \{0,1,2,\dots\}$ and 
a value $K\in\{0,1,2,\dots\}$, decide whether $\nu^{max}_B(G)\leq K$?
\end{problem}

\begin{problem}[Decision version of network flow security]\label{prob:nfs_decision}
Given an interdiction network $G$ and an interdiction budget $B\in\{0,1,2,\dots\}$,
decide whether $\nu^{\max}_B(G)<\nu^{\max}(G)$?
\end{problem}

It is easy to observe that Problem~\ref{prob:nfs_decision} is a special case of
Problem~\ref{prob:nip_decision} by choosing $K=\nu^{\max}(G)$. Otherwise, when working
on a class of interdiction networks with a single source or sink,
Problem~\ref{prob:nip_decision} can be reduced to Problem~\ref{prob:nfs_decision}
by the following simple construction. Suppose we have a single source $s$ (the case
of a single sink is analogue). We introduce a new vertex $s'$ which replaces $s$ as
source and add an unremovable arc from $s'$ to $s$ with capacity equal to $K$.
Problem~\ref{prob:nfs_decision} on the modified interdiction network is then equivalent
to Problem~\ref{prob:nip_decision} on the initial interdiction network.

The following theorem was shown in \cite{wood_1993_deterministic} by reducing a
maximum clique problem to Problem~\ref{prob:nip_decision}.
\begin{theorem}[\cite{wood_1993_deterministic}]\label{thm:strong_np_general}
Problem~\ref{prob:nip_decision} is strongly NP-complete even when the underlying
interdiction network is restricted to unit interdiction costs.
\end{theorem}

Furthermore there is a trivial reduction from the binary knapsack problem
(see \cite{garey_1979_computers} for more information on the binary knapsack problem)
to an interdiction problem on a graph with only two vertices
\cite{wood_1993_deterministic} implying the following theorem.

\begin{theorem}[\cite{wood_1993_deterministic}]\label{thm:weak_np_planar}
Problem~\ref{prob:nip_decision} is weakly NP-complete on planar graphs even
when restricted to a single source and sink.
\end{theorem}

A pseudo-polynomial algorithm for network interdiction problems on planar graphs
with a single source and sink \cite{phillips_1993_network} show that that this class of
problems is not strongly NP-complete.

By the reducibility of Problem~\ref{prob:nip_decision} to Problem~\ref{prob:nfs_decision},
Theorem~\ref{thm:strong_np_general} and Theorem~\ref{thm:weak_np_planar} apply also for
Problem~\ref{prob:nfs_decision} when working on interdiction networks with a single
source or a single sink.

It is not known whether the class of interdiction problems on planar graphs with multiple sources
and sinks is strongly NP-complete. Furthermore,
the direct reduction from Problem~\ref{prob:nip_decision} to Problem~\ref{prob:nfs_decision} is not possible anymore
on this class of networks. We will introduce in
Section~\ref{sec:pseudopol_nfs_planar} a pseudo-polynomial algorithm for Problem~\ref{prob:nfs_decision}
on the class of planar interdiction networks with multiple sources and sinks and such that
the sum of the demands is equal to the sum of the supplies and equal to the maximum flow
in the initial network. This algorithm does not seem to generalize in a simple
way to Problem~\ref{prob:nip_decision}.

\subsection{Relation between planar network interdiction and
the $k$-densest subgraph problem in planar graphs}

We will show that finding dense subgraphs of a given size on planar graphs can
easily be modelled as a planar network interdiction problem. The problem of
finding a densest subgraphs of size $k$ is often called the $k$-densest subgraph
problem or the $k$-clustering problem and is formally defined as follows. 

\begin{problem}[$k$-densest subgraph problem]\label{prob:k_densest_subgraph}
Given an undirected graph $G=(V,E)$, find a subgraph over $k$ vertices
with a maximum number of edges.
\end{problem}

Whereas Problem~\ref{prob:k_densest_subgraph} is known to be NP-complete on a
wide variety of graph classes \cite{corneil_1984_clustering}, its complexity
for the class of planar graphs is still open. A slight modification of
Problem~\ref{prob:k_densest_subgraph} obtained by imposing that the subgraph
must be connected was shown to be NP-complete on planar graphs \cite{keil_1991_complexity}.

\begin{theorem}\label{thm:kdense}
The $k$-densest subgraph problem on planar graphs can be reduced in polynomial
time to a network interdiction problem on planar graphs. 
\end{theorem}

\begin{proof}
Let $G=(V,E)$ be a planar undirected graph. Consider the following planar interdiction
network $G'=(V',E',u,S,T,c,d)$. The underlying graph $(V',E')$ is obtained from
$G$ by subdividing all edges, i.e., on every edge $e\in E$, a new node
$v_e$ is added. We thus obtain a bipartite planar graph where each edge has one
endpoint in $V$ and the other one in the set of newly added vertices $V_E$ ($V'=V\cup V_E$).
By directing all edges 
from $V$ to $V_E$, we get $E'$ (c.f. Figure~\ref{fig:kdense}). The sets containing the sources
and sinks are defined as follows $S=V$, $T=V_E$, all arcs have unit capacity
$u(e)=1 \;\forall\:e\in E'$, sources have infinite supply $d(s)=\infty \;\forall\: s\in S$
and every sink has unit demand $d(t)=1 \;\forall\: t\in T$. Furthermore all arcs and all vertices
of $V_E$ are unremovable (they have an interdiction cost of $\infty$) and the vertices in
$V$ have an interdiction cost equal to one.

\begin{figure}[t]
\begin{center}
\includegraphics{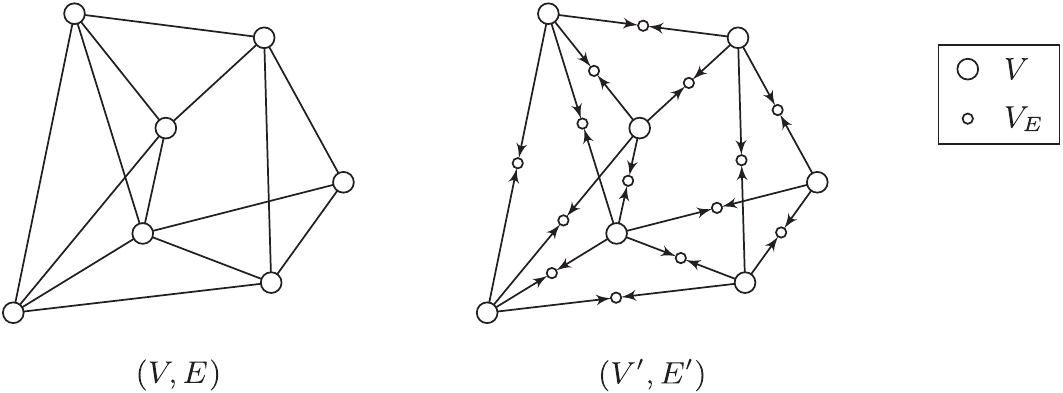}
\end{center}
\caption{Topology of the auxiliary graph $(V',E')$ used for proofing Theorem~\ref{thm:kdense}.}
\label{fig:kdense}
\end{figure}

For some fixed budget $B\in \{0,1,2\dots\}$, an optimal interdiction
set $G'$ corresponds exactly to the vertices of a $B$-densest subgraph in $G$ because of the following
observation.
For some interdiction set $R$, the decrease of flow by removing the components in $R$ corresponds
to the number of sinks for which both neighbors are in $R$. This corresponds to the number of edges
in $G$ that have both endpoints in $R$. 
\end{proof}

\section{Planar duality and current pseudo-polynomial algorithms}\label{sec:duality_and_algos}

Planarity is a very helpful property when dealing with interdiction problems as
the problem seems to have a simpler form when restated on the planar dual of
the original interdiction network. We first introduce
the planar dual of an interdiction network, which can be seen as a generalization of the
classical planar dual. In a second step we propose a pseudo-polynomial algorithm for
planar network interdiction with a single source and a single sink and without vertex
removals. This algorithm is a direct generalization of an algorithm introduced
in \cite{phillips_1993_network}, which was designed only for undirected networks
\footnote{Furthermore the algorithm we present does not allow partial arc
removals whereas the algorithm presented in \cite{phillips_1993_network} did
allow it. This makes no real difference as the technique applies easily to
both cases.}. The extensions we propose in the following sections
will overcome some restrictions of this algorithm.

\subsection{Planar duality for interdiction networks}\label{subsec:planar_duality}

The classical planar dual\footnote{It is also called \emph{geometric dual} or simply \emph{dual}.}
of a directed graph is constructed on the base of a planar embedding
by placing a vertex in each face of the original graph and connecting
two vertices by an arc if they correspond to faces in the original graph sharing an arc.
This gives a natural one-to-one correspondence between arcs in the original graph and
arcs in the dual graph (dual arcs) as well as faces in the original graph and vertices
in the dual graph, and vice versa. By convention, the dual arcs are oriented such that they cross the
corresponding original arcs from right to left. See \cite{lawler_1976_combinatorial} for more details. 

\begin{figure}[t]
\begin{center}
\includegraphics{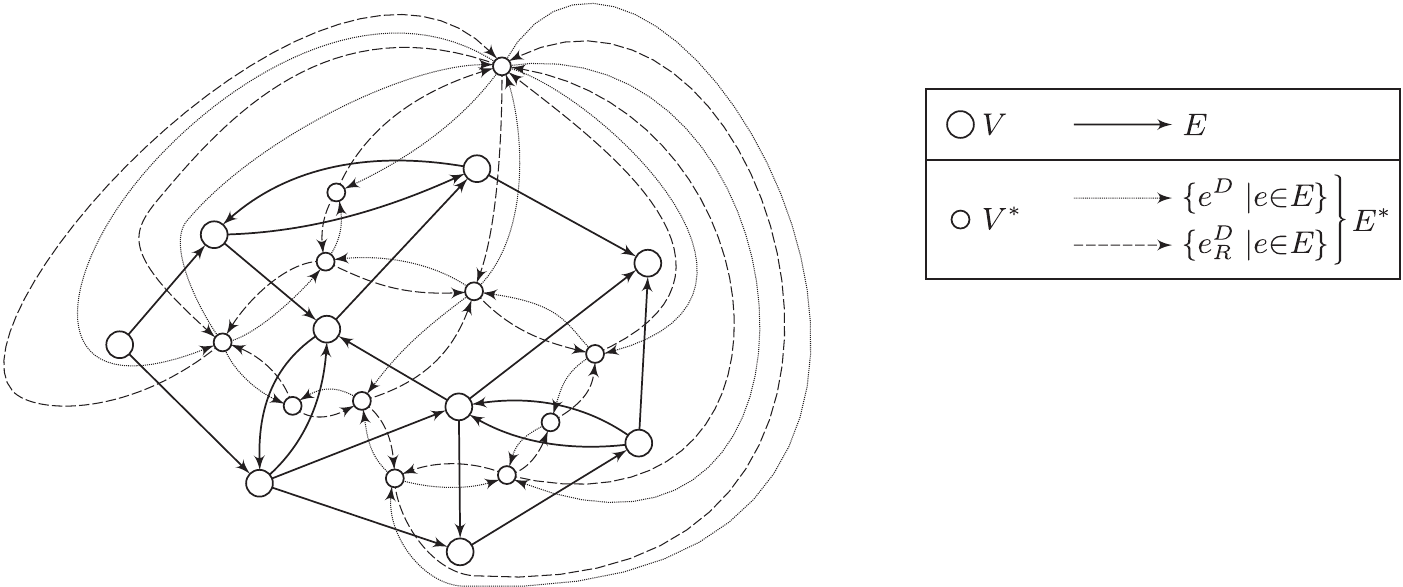}
\end{center}
\caption{Example dual graph $(V^*,E^*)$ drawn over the given original graph $(V,E)$.}
\label{fig:flowDual}
\end{figure}

For our purposes we extend the notion of planar duality on networks with lower and upper capacities
on the arcs and interdiction costs on the arcs. Let $G=(V,E,l,u,c)$ be a directed planar network
where for every arc $e\in E$, $l(e),u(e),c(e) \in \{0,1,2,\dots\}$ correspond to the lower capacity
bound, upper capacity bound and interdiction cost of arc $e$ (where $l(e)\leq u(e) \;\forall\: e\in E$).
We define the dual $G^*=(V^*,E^*,\lambda^*,c^*)$ of the network $G$ in the following way.
The graph $(V^*,E^*)$ is the planar dual in the classical sense of the graph $(V,E)$ with the
single difference that for every arc in the dual we added a reverse arc. For every arc
$e\in E$ we denote by $e^D$ the corresponding dual arc (as in the classical sense) and by
$e^D_R$ its reverse arc (cf. Figure~\ref{fig:flowDual}). The function $\lambda^*:E^*\rightarrow \mathbb{Z}$
is an integral length function in the network $G^*$, defined
by $\lambda^*(e^D)=u(e)$ and $\lambda^*(e^D_R)=-l(e)$ $\forall\: e\in E$. The cost function $c^*$ is
defined by $c^*(e^D)=c(e), c^*(e^D_R)=0$ $\forall\: e\in E$.

For every cut $[V',V\setminus V']$ in the original network we denote its corresponding
dual arcs by $\mathcal{C}^*(V')=\{e^D\in E^*\mid e\in (V',V\setminus V')\}\cup \{e^D_R \in E^*\mid e\in (V\setminus V',V')\}$.
Note that the set $\mathcal{C}^*(V')$ is a set of edge-disjoint,
counterclockwise non-overlapping circuits in $(V^*,E^*)$, where non-overlapping is defined as follows. Let
$\mathcal{C}^*_1,\mathcal{C}^*_2$ be two circuits in $G^*$ and $V_1,V_2\subset V$ be the vertices in $V$ surrounded in
counterclockwise sense by $\mathcal{C}^*_1$ respectively $\mathcal{C}^*_2$. We say that $\mathcal{C}^*_1,\mathcal{C}^*_2$ do
not overlap if $V_1 \cap V_2 = \emptyset$.
The following proposition highlights the correspondence of minimal cuts in the network $G$
and sets of non-overlapping, edge-disjoint circuits in its dual $G^*$.

\begin{proposition}\label{proposition:cuts_to_dual}
The function that associates with every cut $[V',V\setminus V']$ its corresponding dual arcs $\mathcal{C}^*(V')$
is a one-to-one mapping between cuts in $G$ and sets of non-overlapping, edge-disjoint, counterclockwise circuits
in $G^*$.
Furthermore, the value of a cut in $G$ is equal to the sum of the lengths of the corresponding dual
arcs in $G^D$, i.e., for any subset $V' \subset V$, we have
\begin{equation*}
\nu([V',V\setminus V'])=\sum_{e^*\in \mathcal{C}^*(V')}\lambda^*(e^*)\;.
\end{equation*}
\end{proposition}

\begin{proof}
The one-to-one property follows easily by observing that for any set $\mathcal{C}^*_0$ of non-overlapping,
edge-disjoint, counterclockwise circuits, the set $V'$ of all vertices being surrounded in counterclockwise
sense by one of the circuits in $\mathcal{C}^*_0$
satisfies $\mathcal{C}^*(V')=\mathcal{C}^*_0$. The equality between the value of a cut in $G$ and the sum of
the lengths of the corresponding dual arcs follows directly from the definition of $\lambda^*$.
\end{proof}

Proposition~\ref{proposition:cuts_to_dual} implies that for every minimal cut in $G$, there
exists a corresponding circuit in the dual $G^*$ with length equal to the value of the cut. In particular,
when working on a flow network $G=(V,E,l,u,s,t)$ with a single source $s$ and a single
sink $t$, every minimal $s$-$t$ cut in $G$ corresponds to a counterclockwise
circuit around $s$ with $t$ separated from $s$ by the circuit and with length equal to the value of the cut.
We call circuits that separate $s$ and $t$ in such a way \emph{$s$-$t$ separating counterclockwise circuits}.
In the following we see how this correspondence can be exploited for solving network interdiction problems
on planar graphs with a single source and a single sink.

\subsection{A pseudo-polynomial algorithm for single source, single sink
network interdiction on planar graphs without vertex removal}\label{sec:simple_st_interdiction}

We now construct a pseudo-polynomial algorithm for solving the network interdiction problem on
planar directed graphs with a single source $s$ and a single sink $t$ and without vertex removal,
which is a direct generalization of 
an algorithm presented in \cite{phillips_1993_network} (that only worked for undirected graphs).
This algorithm nicely illustrates the techniques currently used for creating pseudo-polynomial network interdiction
algorithms on planar graphs. Given is an interdiction network $G=(V,E,u,s,t,c)$ with unremovable
vertices, i.e., $c(r)=\infty\;\forall\:r\in V$, and interdiction budget $B$. For every interdiction
set $R\subset E$ we fix a minimum $s$-$t$ cut in $G\setminus R$ that we denote by $[V_R,V\setminus V_R]$.
We therefore have 
$\nu^{\max}(G\setminus R)=\nu([V_R,V\setminus V_R])-\sum_{e\in R \cap \omega^+(V_R)}c(e)$.
Note that an optimal, minimal interdiction set $R$ always satisfies $R\subset \omega^+(V_R)$, as otherwise
the interdiction set $R'=R\cap \omega^+(V_R)$ would reduce the maximum flow by the same value as
$R$ and has lower interdiction cost. The reduced value of an $s$-$t$ cut $C$ (with respect to the budget
$B$) is defined as the minimum value of $C$ in $G \setminus R$ over all interdiction sets $R$ and is denoted
by $\nu_{B}(C)$.
Note that the problem to find for some given $s$-$t$ cut $C$ an interdiction
set $R$ that minimizes the value of $C$ in $G\setminus R$ is a binary knapsack problem.

The main idea of the algorithm is to find an optimal, minimum interdiction set $R$ by
finding a corresponding $s$-$t$ cut with minimal reduced value. This is done by translating
the problem into the dual. For any set of edges $U^*\subset E^*$, we define its reduced
length (with respect to $B$) by
$\lambda^*_B(U^*)=\min\{\sum_{e^*\in U^*\setminus X^*}\lambda^*(e^*)
\mid X^* \subset U^*, \sum_{e^*\in X^*}c^*(e^*)\leq B\}$. Similarly for a walk $W^*$ in $G^*$
going along the arcs $(e^*_1,e^*_2,\dots,e^*_k)$, we define
$\lambda^*_B(W^*)=\min\{\sum_{i\in\{1,2,\dots,k\}\setminus I} \lambda^*(e^*_i)\mid I\subset \{1,2,\dots,k\},
\sum_{i\in I}c^*(e^*)\leq B\}$.
By the correspondence between $s$-$t$ cuts in $G$ and $s$-$t$ separating counterclockwise circuits
in $G^*$ as highlighted in Section~\ref{subsec:planar_duality}, we have that
the problem of finding an $s$-$t$ cut in $G$ with minimal reduced value is equivalent to
finding an $s$-$t$ separating counterclockwise circuit with minimal reduced length in the dual.
Such circuits can be described in the following way. Let $P$ be any path in the graph $G$ from vertex $s$ to
vertex $t$, we define $P^D=\{e^D \in E^*\mid e\in P\}$ and $P^D_R=\{e^D_R \in E^*\mid e\in P\}$. For
any set of edges $U^*\subset E^*$ we define its parity with respect to $P$ by
$p_P(U^*)=|U^*\cap P^D|-|U^*\cap P^D_R|$.
By a result of $\cite{miller_1995_flow}$ we have that for every circuit $\mathcal{C}^*$ in $G^*$,
two consecutive crossings on $P$ alternate between left-right crossing and right-left ones. This
implies that every circuit $\mathcal{C}^*$ in $G$ satisfies $p_P(\mathcal{C}^*)\in \{-1,0,1\}$. 
Another implication is that
a circuit $\mathcal{C}^*$ in $G^*$ has the properties to be counterclockwise $s$-$t$ separating
if and only if $p_P(\mathcal{C}^*)=1$.

We therefore have to solve the following problem.

\begin{problem}\label{prob:st_circuit}
\begin{equation*}
\argmin\{\lambda^*_B(\mathcal{C}^*)\mid \mathcal{C}^* \text{circuit in } G^* \text{with } p_P(\mathcal{C}^*)=1\}
\end{equation*}
\end{problem}

Consider the following relaxation of Problem~\ref{prob:st_circuit}.

\begin{problem}\label{prob:st_closedwalk}
\begin{equation*}
\argmin\{\lambda^*_B(W^*)\mid W^* \text{closed walk in } G^* \text{with } p_P(W^*)=1\}
\end{equation*}
\end{problem}

A solution to Problem~\ref{prob:st_circuit} can be easily obtained on the base of a solution
$W^*$ to Problem~\ref{prob:st_closedwalk} by the following observation.
$W^*$ can be partitioned in a disjoint union of circuits
$\mathcal{C}^*_1,\mathcal{C}^*_2,\dots,\mathcal{C}^*_k$. Furthermore, by modularity of the parity
function $p_P$ and the fact that $p_P(W^*)=1$, we have
$1=p_P(W^*)=\sum_{i=1}^k p_P(\mathcal{C}^*_i)$. As the parity of each circuit is
in $\{-1,0,1\}$ we have that there is some index $i\in\{1,2,\dots,k\}$ with $p_P(\mathcal{C}^*_i)=1$.
From $\mathcal{C}^*_i \subset W^*$ follows that $\lambda^*_B(\mathcal{C}^*_i)\leq \lambda^*_B(W^*)$.
By optimality of $W^*$ for Problem~\ref{prob:st_closedwalk} we thus have
$\lambda^*_B(\mathcal{C}^*_i)=\lambda^*_B(W^*)$ and by the fact that
Problem~\ref{prob:st_closedwalk} is a relaxation of Problem~\ref{prob:st_circuit} follows
that $\mathcal{C}^*_i$ is an optimal solution for Problem~\ref{prob:st_circuit}. More generally,
the above reasoning shows that minimal solutions of Problem~\ref{prob:st_closedwalk}
correspond to solutions of Problem~\ref{prob:st_circuit} and vice versa.

We finally show how Problem~\ref{prob:st_closedwalk} can be solved by a dynamic programming approach
that we realize as a sequence of shortest
path problems on an auxiliary graph $G'=(V',E',\lambda')$ with positive arc lengths $\lambda'$
which is defined as follows. Let $P$ be a shortest path from $s$ to $t$ in $G$ and we denote by
$|P|$ the length of the path which is the number of arcs used in $P$.
$V'$ consists of $|V^*|\cdot (B+1)\cdot (2|P|+1)$ vertices that we
denote in the following way $V'=\{v^*_{b,p} \mid v^*\in V^*, b\in\{0,1,\dots,B\}, p\in\{-|P|,-|P|+1,\dots,|P|\}$.
The set of arcs $E'$ is defined by $E'=E'_0\cup E'_{1,=}\cup E'_{1,+}\cup E'_{1,-}\cup E'_{2,=}\cup E'_{2,+}\cup E'_{2,-}$ with
\begin{compactitem}
\item $E'_0=\{(v^*_{b,p},v^*_{b-1,p})\mid v^*_{b,p}\in V', b\in\{1,2,\dots,B\}, p\in\{-|P|,-|P|+1,\dots,|P|\} \}$
\item $E'_{1,=}=\{(u^*_{b,p},v^*_{b,p})\mid (u^*,v^*)\in E^*\setminus (P^D\cup P^D_R),
        b\in\{0,1,\dots,B\},p\in\{-|P|,-|P|+1,\dots,|P|\} \}$
\item $E'_{1,+}=\{(u^*_{b,p},v^*_{b,p+1})\mid (u^*,v^*)\in P^D,
        b\in\{0,1,\dots,B\},p\in\{-|P|,-|P|+1,\dots,|P|-1\} \}$
\item $E'_{1,-}=\{(u^*_{b,p},v^*_{b,p-1})\mid (u^*,v^*)\in P^D_R,
        b\in\{0,1,\dots,B\},p\in\{-|P|+1,-|P|+2,\dots,|P|\} \}$
\item $E'_{2,=}=\{(u^*_{b_1,p},v^*_{b_2,p})\mid (u^*,v^*)\in E^*\setminus (P^D\cup P^D_R),
        b_1\in\{c^*(u^*,v^*),c^*(u^*,v^*)+1,\dots,B\}, b_2=b_1-c^*(u^*,v^*),p\in\{-|P|,-|P|+1,\dots,|P|\} \}$
\item $E'_{2,+}=\{(u^*_{b_1,p},v^*_{b_2,p+1})\mid (u^*,v^*)\in P^D,
        b_1\in\{c^*(u^*,v^*),c^*(u^*,v^*)+1,\dots,B\}, b_2=b_1-c^*(u^*,v^*),p\in\{-|P|,-|P|+1,\dots,|P|-1\} \}$
\item $E'_{2,-}=\{(u^*_{b_1,p},v^*_{b_2,p-1})\mid (u^*,v^*)\in P^D_R,
        b_1\in\{c^*(u^*,v^*),c^*(u^*,v^*)+1,\dots,B\}, b_2=b_1-c^*(u^*,v^*),p\in\{-|P|+1,-|P|+2,\dots,|P|\} \}$.
\end{compactitem}

We define a function $\eta:V'\cup(E'\setminus E'_0)\rightarrow V\cup E$ that maps elements of
$G'$ to corresponding elements in $G^*$ in the following way.
$$\begin{array}{rll}
\eta(v^*_{b,p})&=v^* &\forall v^*_{b,p} \in V'\\
\eta(v^*_{b_1,p_1},u^*_{b_2,p_2})&=(v^*,u^*) &\forall (v^*_{b_1,p_1},u^*_{b_2,p_2})\in E'\setminus E'_0
\end{array}$$

The length $\lambda':E'\rightarrow \{0,1,\dots\}$ is defined as follows.
\begin{equation*}
\lambda'(e')=\begin{cases}
0 & \text{if } e'\in E'_0\cup E'_{2,=}\cup E'_{2,+}\cup E'_{2,-}\\
\lambda^*(\eta(e')) & \text{if } e'\in E'_{1,=}\cup E'_{1,+}\cup E'_{1,-}
\end{cases} 
\end{equation*}

We now define a correspondence between walks in $G'$ and walks in $G^*$ by extending the function $\eta$ in the following way.
Let $u^*_{b_1,p_1},v^*_{b_2,p_2}\in V'$, $W'$ be a walk in $G'$ from $u^*_{b_1,p_1}$ to $v^*_{b_2,p_2}$ and $(e'_1,e'_2,\dots,e'_k)$
the suite of edges corresponding to the walk $W'$. Furthermore, let $I=\{i\in\{1,2,\dots,k\}\mid e'_i \in E'\setminus E_0\}$.
The walk $\eta(W')$ in $G'$ is defined to go along the edges $(\eta(e'_i))_{i\in I}$. It is easy to verify that $\eta(W')$ is
effectively a walk (from $u^*$ to $v^*$) in $G^*$ and satisfies $p_P(\eta(W'))=p_2-p_1$. We associate to the walk
$W'$ a set $R(W')\subset E$ defined by $R(W')=\{e\in E\mid \exists i\in I \text{ with } e'_i\in E'_{2,=}\cup E'_{2,+}\cup E'_{2,-}
\text{ and } e^D=\eta(e'_i) \}$. By construction of $G'$ we have $c(R(W'))\leq b_1-b_2$ \footnote{When $\eta(W')$ corresponds to a path
or a circuit, this relation is always satisfied with equality.}. Therefore, for any path $W'$ in $G'$, the set $R(W')$
is an interdiction set. Furthermore, we have $\lambda^*_{b_1-b_2}(\eta(W'))\leq \lambda'(W')$ because by setting the length of
the arcs $R(W')^D$ in $G^*$ to zero, the length of $\eta(W')$ is smaller or equal than $\lambda'(W')$ \footnote{Here too, in case
that $\eta(W')$ corresponds to a path or a circuit, we have $\lambda^*_{b_1-b_2}(\eta(W'))=\lambda'(W')$.}. 

In particular, if the walk $W'$ goes from $v^*_{B,0}$ to $v^*_{0,1}$ for some $v^*\in V^*$, we then have that $\eta(W')$ is
a closed walk in $G^*$ with $p_P(\eta(W'))=1$ and satisfying $\lambda^*_B(\eta(W'))\leq \lambda'(W')$. Conversely, for every
circuit $C^*$ in $G^*$ satisfying $p_P(C^*)=1$, we can find a corresponding path $P'$ in $G'$ with $\lambda'(P')=\lambda^*_B(C^*)$
in the following way. Let $(e^*_1,e^*_2,\dots,e^*_k)$ be a suite of arcs in $E^*$ corresponding to the circuit $C^*$ and $v^*\in V^*$ be
the vertex corresponding to the tail of $e^*_1$. Let $I\subset\{1,2,\dots,k\}$ be a set satisfying $\sum_{i\in I}c^*(e^*_i)\leq B$ and
$\lambda^*_B(C^*)=\sum_{i\in I}\lambda^*(e^*_i)$.

The path $P'$ will be defined by its corresponding suite of arcs
$(e'_1,e'_2,\dots,e'_{k'})$ as follows. We set $k'=k+B-\sum_{i\in I}c^*(e^*_i)$.
$P'$ starts at the vertex $v^*_{B,0}$. The path $P'$ will be defined by adding edges step
by step. For $i\in \{1,2,\dots,k\}$ let $e^*_i=(u^*,v^*)$ and
$u^*_{b,p}$ be the endpoint of the currently constructed path. If $i\in I$, we set
$e'_i=(u^*_{b,p},v^*_{b-c^*(e^*_i),p+p_P(e^*_i)})$ and otherwise $e'_i=(u^*_{b,p},v^*_{b,p+p_P(e^*_i)})$.
For $i\in \{k+1,k+2,\dots,k'\}$, we set $e'_i=(v^*_{k'-i+1,1},v^*_{k'-i,1})$. It is easy to verify
that $P'$ is effectively a path in $G'$ with $\eta(P')=\mathcal{C}^*$ satisfying $\lambda'(P')=\lambda^*_B(C^*)$.

Let $v^*\in V^*$ be some fixed vertex and $P'$ be a shortest path from $v^*_{B,0}$ to $v^*_{0,1}$.
The discussion above shows that $\eta(P')$ is a solution to the following problem.
\begin{problem}\label{prob:st_closedwalk_containing_v}
\begin{equation*}
\argmin\{\lambda^*_B(W^*)\mid W^* \text{closed walk in } G^* \text{containing vertex } v^* \text{and satisfying } p_P(W^*)=1\}
\end{equation*}
\end{problem}

Solving Problem~\ref{prob:st_closedwalk_containing_v} for all vertices $v^*\in V^*$ will solve
Problem~\ref{prob:st_closedwalk}. However, as a walk solving Problem~\ref{prob:st_closedwalk}
will pass at least once by an arc $e'$ with $\eta(e')\in P^D$ it suffices to solve
Problem~$\ref{prob:st_closedwalk_containing_v}$ for all vertices in $G^*$ having at least
one outgoing edge contained in $P^D$. Therefore, it suffices to solve $|P|$ instances of
Problem~\ref{prob:st_closedwalk_containing_v} to get a solution to
Problem~\ref{prob:st_closedwalk}. Notice that solving the network interdiction problem
with the above algorithm for some budget $B$, solves also the interdiction problems
for all budgets $B'<B$. This solution can be obtained in an analogue manner as for
the case with budget $B$ by looking in $G'$ at walks from $v^*_{B,0}$ to $v^*_{B-B',1}$.

Applying Dijkstra's algorithm for shortest paths to
solve the subproblems of Problem~\ref{prob:st_closedwalk_containing_v}, the algorithm introduced
above to solve Problem~\ref{prob:st_closedwalk} gets an overall complexity of
$\mathcal{O}(B |P|^2 n \log(n B))$. Because of the special structure of the network $G'$
it is easy slightly improve the running time by handling the vertices in Dijkstra's algorithm
level by level with respect to the remaining budget, i.e., we handle 
all vertices corresponding to a budget $B$ first (always taking the one with the smallest
label as usual), then all with budget $B-1$ 
and so on. When looking for the vertex with the smallest label, we only have $O(n)$
candidates instead of $O(n B)$. This allows to reduce the running time to
$\mathcal{O}(B |P|^2 n \log(n))$. By reversing the roles of budget and length in the proposed algorithm,
one can replace $B$ in the above running times by $\nu_B^{\max}(G)$.  

Notice that the proposed method can be slightly simplified (without
influencing the above running time bounds) allowing
for the parity only to take values in
$\{-\lceil\frac{|P|}{2}\rceil, -\lceil\frac{|P|}{2}\rceil+1, \dots, \lceil\frac{|P|}{2}\rceil\}$ instead
of $\{-|P|,-|P|+1,\dots,|P|\}$ as every circuit in $G^*$ with parity equal to one contains no subpath with
a parity not contained in this range because it would not be possible to come back to parity one without
going two times through a same vertex.

The formulation of Problem~\ref{prob:st_closedwalk_containing_v} as a shortest path
problem on $G'$ is essentially a dynamic programming realization of a multi-objective shortest path problem
with the three objectives budget, length and parity on a graph $\overline{G}$ defined as follows.
The graph $\overline{G}=(V^*,\overline{E})$ is obtained from the graph $G^*=(V^*,E^*)$ by doubling every arc.
For every arc $e^D\in E^*$, we denote by $\overline{e}_1,\overline{e}_2$ the two corresponding
parallel arcs in $\overline{E}$. To every arc in $\overline{E}$ we associate 
a parity value, a length and a budget value. The parity value
of $\overline{e}_1$ and $\overline{e}_2$ is set to $p_P(e)$. The length of $\overline{e}_1$ is
equal to $u(e)$ and the length of $\overline{e}_2$ is set to zero. Finally, the budget value is
set to zero for $\overline{e}_1$ and to $c(e)$ for $\overline{e}_2$. Therefore, the arcs in $\overline{E}$
indexed by one correspond to non-removed arcs and the ones indexed by two correspond to removed arcs. For some
fixed vertex $v^*\in V^*$, a closed walk in $\overline{G}$ containing $v$, having parity equal to one, a budget
value bounded by $B$ and with minimal length among all those closed walks corresponds exactly to a solution
to Problem~\ref{prob:st_closedwalk_containing_v}.

\section{Incorporating vertex interdiction and vertex capacities}\label{sec:vertex_interdiction}

Let $G=(V,E,u,s,t,c)$ be an interdiction network. In this section we show how vertex interdiction
and vertex capacities can be incorporated in the algorithm of
the previous section. We begin by introducing the possibility of vertex interdiction and
observe afterwards how vertex capacities can be added to the model.
The role of the dual network $G^*$ will be replaced by a modified dual $\widetilde{G^*}$ which allows
to model vertex interdiction basically as arc interdiction.
This technique was used in \cite{khuller_1994_flow} for modelling vertex capacities in planar
flow problems. As we will see in this section, this model can also be used for modelling
vertex interdiction. The modified dual we present in this section and extend in the next one
is slightly different from the one presented in $\cite{khuller_1994_flow}$, using some less
artificially added vertices.

\begin{figure}[t]
\begin{center}
\includegraphics{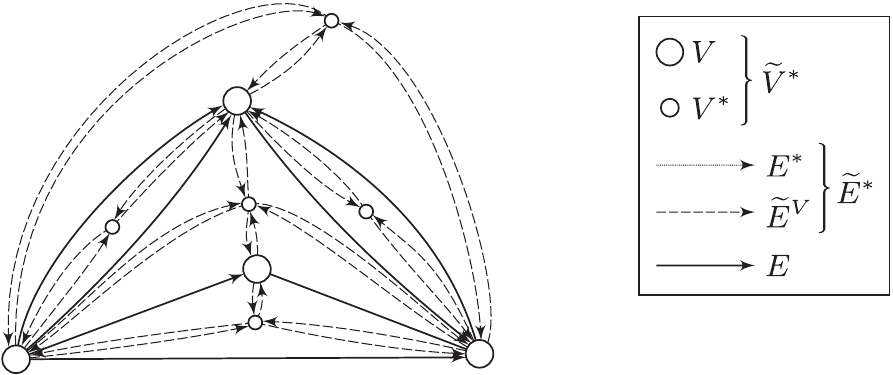}
\end{center}
\caption{Topology of the auxiliary graph $(\widetilde{V}^*,\widetilde{E}^*)$ used for modeling vertex interdiction.}
\label{fig:vertexInterdiction}
\end{figure}

The modified dual network
$\widetilde{G^*}=(\widetilde{V^*},\widetilde{E^*},\widetilde{\lambda^*},\widetilde{c^*})$ is an
extended version of $G^*=(V^*,E^*,\lambda^*,c^*)$
with additional vertices and arcs defined as follows. For every vertex $v\in V$ we denote by $f^*(v)$ the
face of $G^*$ corresponding to $v$. Similarly, for every $v^*\in V^*$ we denote by $f(v^*)$ the
face of $G$ corresponding to $v^*$. 
The network $\widetilde{G^*}$ consists of the vertices $V^*$ and contains an additional vertex for every face
of $G^*$. Because of the one-to-one relation between faces in $G^*$ and vertices in $G$ we set
$\widetilde{V^*}=V^* \cup V$. The arc set $\widetilde{E^*}$ is defined by
$\widetilde{E^*}=E^* \cup \widetilde{E^V}$ where $\widetilde{E^V}$ contains the two arcs
$(v,v^*)$ and $(v^*,v)$
for every pair of $v\in V$ and $v^*\in V^*$, where $f(v^*)$ is a face adjacent to $v$
(c.f. Figure~\ref{fig:vertexInterdiction}).
The length function $\widetilde{\lambda^*}$ is an extension of $\lambda^*$ on the
arcs $\widetilde{E^*}$ giving a value of $\infty$ to all arcs in $\widetilde{E^V}$ (or simply a high finite value
assuring that the arc is never used in following applications of shortest path algorithms).
Furthermore, the modified interdiction cost function $\widetilde{c^*}$ is an extension of $c^*$
on the arcs in $\widetilde{E^*}$ defined as follows.
$$\begin{array}{rll}
\widetilde{c^*}(e^*)&=c^*(e^*) &\forall e^* \in E^*\\
\widetilde{c^*}(v,v^*)&=0 &\forall (v,v^*)\in \widetilde{E^V}\cap V \times V^* \\
\widetilde{c^*}(v^*,v)&=c(v) &\forall (v^*,v)\in \widetilde{E^V}\cap V^* \times V
\end{array}$$
For some subset of arcs $\widetilde{U^*}\subset \widetilde{E^*}$ we define their reduced cost in $\widetilde{G}^*$
with respect to the budget $B$ by 
$\widetilde{\lambda^*_B}(\widetilde{U^*}):=\min\{\sum_{\widetilde{e^*}\in \widetilde{U^*}%
\setminus X^*} \widetilde{\lambda^*}(\widetilde{e^*})%
\mid X^* \subset \widetilde{U^*}, \sum_{\widetilde{e^*}\in X^*}\widetilde{c^*}(\widetilde{e^*})\leq B\}$.
The main idea of this model lies in the following correspondence between interdiction
sets in $G$ and $s$-$t$ separating counterclockwise circuits in $\widetilde{G^*}$.

\begin{theorem}\label{thm:vertex_interdiction}
\hspace{0mm}\\[-\baselineskip]
\begin{enumerate}[i) ]
\item For every interdiction set $R\subset V\cup E$ with respect to the budget $B$, there
is an $s$-$t$ separating counterclockwise circuit $\widetilde{\mathcal{C}^*}$ in $\widetilde{G^*}$
satisfying $\widetilde{\lambda^*_B}(\widetilde{\mathcal{C}^*})\leq \nu^{\max}(G\setminus R)$.
\item For every $s$-$t$ separating counterclockwise circuit in $\widetilde{G^*}$ and every budget $B$,
there is an interdiction set $R\subset V\cup E$ with respect to $B$ satisfying
$\nu^{max}(G\setminus R)\leq \widetilde{\lambda^*_B}(\widetilde{\mathcal{C}^*})$.
\end{enumerate}
\end{theorem}

\begin{proof}
Let $R\subset V\cup E$ be an interdiction set with respect to some fixed
budget $B$ and $U\subset E$ be a set of arcs corresponding to a minimum $s-t$ cut
in $G\setminus R$. As there is no path from $s$ to $t$ in the graph $G\setminus(R \cup U)$,
it is easy to see (by means of the Jordan curve theorem) that there is an $s$-$t$ separating
counterclockwise circuit $\widetilde{\mathcal{C}^*}$ in $\widetilde{G^*}$
consisting only of arcs which are either adjacent to vertices in $R$ or are dual arcs of arcs contained in
$R$ or $U$. We therefore have as desired $\widetilde{\lambda^*_B}(\widetilde{C}^*)\leq \nu^{\max}(G\setminus R)$.
        
Conversely let $\widetilde{\mathcal{C}^*}$ be
an $s$-$t$ separating counterclockwise circuit in $\widetilde{G^*}$
satisfying $\widetilde{\lambda_B^*}(\widetilde{\mathcal{C}^*})<\infty$ (when the reduced dual
length is equal to $\infty$, the result follows trivially). 
Let $\widetilde{U^*}\subset \widetilde{E^*}$ be a solution of
\begin{equation*}
\underset{\widetilde{X^*}\subset \widetilde{\mathcal{C}^*}}{\argmin}%
\{\sum_{\widetilde{e^*}\in \widetilde{\mathcal{C}^*} \setminus \widetilde{X^*}}\lambda^*(e^*)%
\mid \widetilde{c^*}(\widetilde{X^*})\leq B\}\;.
\end{equation*}

By definition of the reduced dual length we have
$\widetilde{\lambda^*_B}(\widetilde{\mathcal{C}^*})=%
\widetilde{\lambda^*}(\widetilde{\mathcal{C}^*}\setminus \widetilde{U^*})$.
Let $V_{\widetilde{\mathcal{C}^*}}$ be
the subset of vertices in $V$ through which the circuit $\widetilde{\mathcal{C}^*}$ passes
and let $U^*=\widetilde{U^*}\cap E^*$.
Because of $\widetilde{\lambda^*_B}(\widetilde{\mathcal{C}^*})<\infty$ we have that
all arcs of $\widetilde{\mathcal{C}^*}$ entering in one of the vertices in $V_{\widetilde{\mathcal{C}^*}}$
are contained in $\widetilde{U^*}$. The cost of $\widetilde{U^*}$ can therefore be reformulated as follow.
\begin{equation*}
\widetilde{c^*}(\widetilde{U^*})=c^*(U^*)+c(V_{\widetilde{\mathcal{C^*}}})
\end{equation*}
Let $U=\{e\in E \mid e^D \in U^*\}$ and we define $R=V_{\widetilde{\mathcal{C}^*}}\cup U$.
By the above equation and the definition of $\widetilde{U^*}$ we have
$c(R)=c^*(U^*)+c(V_{\widetilde{\mathcal{C^*}}})=\widetilde{c^*}(\widetilde{U^*})\leq B$ showing that
$R$ is an interdiction set with respect to the budget $B$.
Let $E^*_{\widetilde{\mathcal{C}^*}}=\widetilde{\mathcal{C}^*}\cap E^*$ and
$E_{\widetilde{\mathcal{C}^*}}=\{e\in E \mid e^D \in E^*_{\widetilde{\mathcal{C}^*}}\}$. The fact that
$\widetilde{\mathcal{C}^*}$ is a counterclockwise, $s$-$t$ separating circuit implies that there is no path from 
$s$ to $t$ in $G\setminus (V_{\widetilde{\mathcal{C}^*}}\cup E_{\widetilde{\mathcal{C}^*}})$. Therefore,
removing the arcs $E_{\widetilde{\mathcal{C}^*}}\setminus U$ from $G\setminus R$ destroys all paths from
$s$ to $t$ implying $\nu^{\max}(G\setminus R)\leq u(E_{\widetilde{\mathcal{C}^*}}\setminus U)$. The result is
finally obtained by observing that
$u(E_{\widetilde{\mathcal{C}^*}}\setminus U)=\widetilde{\lambda^*}(\widetilde{\mathcal{C}^*}\setminus \widetilde{U^*})=%
\widetilde{\lambda^*_B}(\widetilde{\mathcal{C}^*})$.
\end{proof}

Theorem~\ref{thm:vertex_interdiction} implies that the problem of finding by how much the value of a maximum flow
can be reduced through interdiction reduces to finding an $s$-$t$ separating
counterclockwise circuit with minimal reduced value in $\widetilde{G^*}$. Furthermore, the proof of Theorem~\ref{thm:vertex_interdiction}
shows how one can transform such a circuit to an optimal interdiction set. 

For characterizing $s$-$t$ separating counterclockwise circuits, we introduce an adapted version of the parity function.  
As in the previous section, let $P$ be a path in $G$ from vertex $s$ to vertex $t$. Because in the graph
$\widetilde{G^*}$ it is possible to cross $P$ at a vertex, we have to take this possibility into account in
the parity function. We therefore define a parity function $\widetilde{p_P}$ which is an extension of $p_P$
on the subsets of $\widetilde{E^*}$ in the following way. For every vertex $v\in V\setminus \{s,t\}$ which lies
on the path $P$ and every arc $(v,v^*)\in \widetilde{E^*}$ which leaves $P$ on the left side, we set $\widetilde{p_P}(v,v^*)=1$.
Similarly, for every vertex $v\in V\setminus \{s,t\}$ on $P$ and every arc $(v^*,v)\in \widetilde{E^*}$ which
enters $P$ from the left side we set $\widetilde{p_P}(v^*,v)=-1$. For all other edges in $\widetilde{E^*}$ we set
$\widetilde{p_P}=0$. Finally,
for any set $\widetilde{U^*}\subset \widetilde{E^*}$ we define its parity by $\widetilde{p_P}(\widetilde{U^*})=%
\sum_{\widetilde{e^*}\in \widetilde{U^*}}\widetilde{p_P}(\widetilde{e^*})$.
As in the previous section, we have, by this definition of
$\widetilde{p_P}$, that a circuit $\widetilde{\mathcal{C}^*}$
in $\widetilde{G^*}$ has the properties to be counterclockwise $s$-$t$ separating 
if and only if $\widetilde{p_P}(\widetilde{\mathcal{C}^*})=1$.

Applying the techniques of the previous section to the modified dual $\widetilde{G^*}$ allows therefore
to solve network interdiction problems on planar graphs with a single source and sink and with the
possibility to interdict arcs and vertices. The asymptotic worst case complexity stays the same as
in the previous section because the size of the graph $\widetilde{G^*}$ is only at most a constant factor
larger than $G^*$, and the same is true for the number of times Problem~\ref{prob:st_closedwalk} has to be solved.

Upper capacities on the vertices can be introduced in the same way as shown in \cite{khuller_1994_flow} by slightly
modifying the network $\widetilde{G^*}$ as follows.
Suppose that some vertex $v\in V$ has an upper capacity $u(v)\in \{1,2,\dots\}$. Let 
$V^*_v=\{v^*\in V^*\mid (v^*,v)\in \widetilde{E^*}\}$. For every $v^*\in V^*_v$, an additional arc
from $v^*$ to $v$ is added with interdiction cost $\infty$ and capacity $u(v)$. For a justification
of this construction in the case without interdiction (respectively by setting $B=0$), see \cite{khuller_1994_flow}.
Theorem~\ref{thm:vertex_interdiction} remains true for the modified network modelling vertex capacities. This can
be proven in an analogue way as for the case without vertex capacities. To simplify further results, we consider
in the following only flow networks without vertex capacities. Vertex capacities can easily be added by the above construction.

The method presented in the next section to solve the network flow security problem on networks with multiple
sources and sinks differs significantly from the approach presented in Section~\ref{sec:simple_st_interdiction}.
Therefore the model presented in this section does not apply directly to this case. We will see in the
next section how the model for vertex interdiction can be adapted (the adaption for vertex capacities
will be analogue).

\section{Pseudo-polynomial algorithm for network flow security}\label{sec:pseudopol_nfs_planar}

In this section we propose a pseudo-polynomial algorithm for the network flow security problem on planar
graphs with multiple sources and sinks where the sum of the demands is equal to the sum of the
supplies and equal to the maximum flow. The method is
inspired by an algorithm introduced in \cite{miller_1995_flow} for testing if all demand can be satisfied
in a planar flow problem with multiple sources and sinks. We will
therefore begin with a short summary of this maximum flow algorithm for planar graphs in
Subsection~\ref{subsec:miller_1995_flow}. Our algorithm for network flow security
will be introduced in Subsection~\ref{subsec:netsecalgo} restricted to the case without vertex removals.
This restriction will be lifted in Subsection~\ref{subsec:netsecalgo_vertex_interdiction} by adapting
the modelling idea of Section~\ref{sec:vertex_interdiction}.

\subsection{Maximum flow algorithm of Miller and Naor}\label{subsec:miller_1995_flow}

In this section we briefly present an algorithm introduced in \cite{miller_1995_flow} for testing if
all demands can be satisfied in a flow problem with multiple sources and sinks such that the sum of all demands
equals the sum of all supplies. Let $G=(V,E,l,u,S,T,d)$ be a planar flow network with lower and upper limits on the
capacities designated by $l$ and $u$, with source set $S\subset V$ and sink set $T\subset V\setminus S$
and with demand/supply function $d$ satisfying $-d(S)=d(T)$. A flow in $G$ is called \emph{saturating} if it
satisfies all demands. The problem we want to solve is the following.

\begin{problem}\label{prob:miller}
Does there exist a saturating flow in $G$?
\end{problem}

In a first step, the problem is reduced to a circulation problem on a network
$\widehat{G}$, which 
is the problem of finding a flow in a network with lower and upper capacities on the
arcs but without sources and sinks. The network
$\widehat{G}=(\widehat{V}=V,\widehat{E},\widehat{l},\widehat{u})$ is defined as
follows. Let $\mathcal{T}'$ be an undirected tree over the vertices in $V$ that spans the sources
and sinks and that can be
added to $G$ without destroying planarity. For every edge $\{v,u\}\in\mathcal{T}'$ we denote
by $V_{\mathcal{T}'}(v,u)\subset V$ the set of vertices connected to vertex $v$ in
$\mathcal{T}'\setminus \{v,u\}$. We orient the edges in $\mathcal{T}'$ to obtain
$\mathcal{T}$ in the following way. For $\{v,u\}\in \mathcal{T}'$ we orient the edge from
$v$ to $u$ if $d(V_{\mathcal{T}'}(v,u))\geq 0$, otherwise we orient the edge from $u$ to $v$.
The set $\widehat{E}$ is defined to be $E \cup \mathcal{T}$. Furthermore, the lower and upper capacities
$\widehat{l}$ and $\widehat{u}$ are extensions of $l$ and $u$ on the set $\widehat{E}$ defined by
$\widehat{l}(v,u)=\widehat{u}(v,u)=d(V_{\mathcal{T}'}(v,u))\;\forall\: (v,u)\in\mathcal{T}$.
As noted in \cite{miller_1995_flow} we have the following theorem.
\begin{theorem}\label{thm:reduction_to_circulation}
There exists a saturating flow in $G$ $\Leftrightarrow$ there exists a circulation in $\widehat{G}$.
\end{theorem}
A circulation
in a planar flow network can be computed by solving shortest path problem in its dual in the following way. Let
$\widehat{G}^*=(\widehat{V}^*,\widehat{E}^*,\widehat{\lambda}^*)$ be the dual network
of $\widehat{G}$ as defined in Section~\ref{sec:duality_and_algos}, with the difference that
we have no dual costs $\widehat{c}^*$ as we deal with a standard flow network and not
an interdiction network. Let $\widehat{r}^*\in\widehat{V}^*$ be an arbitrary vertex in
$\widehat{G}^*$ and for $\widehat{v}^*\in\widehat{V}^*$ let $\mu(\widehat{v}^*)$ be
the distance of a shortest path from $\widehat{r}^*$ to $\widehat{v}^*$ (with respect
to the length $\widehat{\lambda}^*$). In \cite{miller_1995_flow} the following theorem
was proven.
\begin{theorem}\label{thm:miller}
There exists a circulation in $\widehat{G}$ $\Leftrightarrow$ $\widehat{G}^*$ has no negative
circuits.
\end{theorem}

In this case, a circulation can
be obtained in the following way. Let $(\widehat{v},\widehat{u})\in \widehat{E}$ and
$(\widehat{v}^*,\widehat{u}^*)=(\widehat{v},\widehat{u})^D$ its corresponding
dual arc. A circulation can finally be defined by assigning
a flow equal to $\max\{0,\mu(\widehat{u}^*)-\mu(\widehat{v}^*)\}$
to each arc $(\widehat{v},\widehat{u})\in \widetilde{E}$.

\subsection{Network security on planar graphs with multiple sources and sinks}\label{subsec:netsecalgo}

Let $G=(V,E,l,u,S,T,d,c)$ be a planar interdiction network satisfying $-d(S)=d(T)$ and without
vertex removal. Let $(\widehat{V},\widehat{E},\widehat{l},\widehat{u})$ be the auxiliary
network corresponding to $(V,E,l,u,S,T,d)$ as defined in Section~\ref{subsec:miller_1995_flow}.
We extend this auxiliary network to
$\widehat{G}=(\widehat{V},\widehat{E},\widehat{l},\widehat{u},\widehat{c})$
where $\widehat{c}$ is an extension of $c$ on the set $\widehat{E}$, satisfying
$\widehat{c}(\widehat{e})=\infty\;\forall\:\widehat{e}\in\widehat{E}\setminus E$.
Let $\widehat{G}^*=(\widehat{V}^*,\widehat{E}^*,\widehat{\lambda}^*,\widehat{c}^*)$ be its
corresponding dual network as defined in Section~\ref{sec:duality_and_algos}.
Using Theorem~\ref{thm:miller} we can formulate a problem on $\widehat{G}^*$,
which is equivalent to Problem~\ref{prob:nfs_decision} for $G$, as follows.
\begin{problem}\label{prob:nfs_decision_dual}
Does there exist a circuit $\widehat{\mathcal{C}}^*$ in $\widehat{G}^*$ 
with $\widehat{\lambda}^*_B(\widehat{\mathcal{C}}^*)<\nu^{\max}(G)$?
\end{problem}

The circuit with minimum reduced cost can be found by similar techniques as
the ones presented in Section~\ref{sec:duality_and_algos} with the differences that we do not
have to take parity into account and that we have
to use a shortest path algorithm, which can deal with negative arc lengths. A simple implementation
would be first to determine shortest paths for all pairs of vertices in $\widehat{G}$. By using an algorithm
presented in \cite{fakcharoenphol_2001_planar} this can be done in $\mathcal{O}(n^2\log^3 n)$ time. 
To check if there is a circuit with negative reduced value in $\widehat{G}^*$
going through some fixed vertex $\widehat{v}^*\in \widehat{V}^*$, a shortest path
in an auxiliary network as described in Section~\ref{sec:duality_and_algos} will be determined.
This can be done by determining the shortest paths from some initial vertex to all others
by proceeding level by level with respect to the remaining budget using at each level information
of the preprocessing step. Over the $B$ budget levels, this can be done in
$\mathcal{O}(B n^2)$ time. By simply performing this operation for all possible start vertices,
we get a running time of $\mathcal{O}(B n^3)$. By reversing the roles of budget and length, we can
get an algorithm with running time $\mathcal{O}(\nu_B^{\max}(G) n^3)$.
We expect that this running time can even be improved by exploiting more deeply the
structure of the auxiliary graph.

\subsection{Generalization to the case with vertex interdiction}\label{subsec:netsecalgo_vertex_interdiction}

Let $G=(V,E,l,u,S,T,d,c)$ be a planar interdiction network satisfying $-d(S)=d(T)$ and allowing arc
and vertex removal (except for sources and sinks).
As in the case without vertex removal, we begin by reformulating the problem
as an interdiction problem for circulations. Let
$\widehat{G}=(\widehat{V}=V,\widehat{E},\widehat{l},\widehat{u},\widehat{c})$ be the auxiliary graph
as defined in Section~\ref{subsec:netsecalgo} and as before we denote by $\mathcal{T}=\widehat{E}\setminus E$
the added tree arcs.
We now discuss how arc and vertex removal in $G$
can be translated to $\widehat{G}$ such that Theorem~\ref{thm:reduction_to_circulation} remains
valid for the resulting networks.
As already exploited in Section~\ref{subsec:netsecalgo}, removing an arc of $G$ corresponds
to removing the same arc in $\widehat{G}$. However, vertex removal cannot be translated in such a
direct way as the arcs in $\widehat{E}\setminus E$ are auxiliary
arcs which should not be removed by a vertex removal.
For any interdiction set
$R\subset V \cup E$, we denote by $\widehat{G}(R)$ the graph obtained from $\widehat{G}$ by removing
all arcs contained in $R$ and all arcs in $E$ being adjacent to a vertex in $R$.
We have the following relation.

\begin{theorem}\label{thm:reduction_to_circulation_ext}
For any interdiction set $R$ of $G$ we have the following equivalence.
\begin{equation*}
\text{There is a saturating flow in $G\setminus R$} \Leftrightarrow%
\text{there is a circulation in $\widehat{G}(R)$.}
\end{equation*}
\end{theorem}

\begin{proof}
Let $G(R)$ be the network obtained from $G$ by removing all arcs in $R$ and all arcs
adjacent to vertices in $R$. We trivially have that there is a saturating flow
in $G\setminus R$ if and only if there is a saturating flow in $G(R)$. The network
$\widehat{G}(R)$ can easily be obtained from $G(R)$ by applying the construction
introduced in Section~\ref{subsec:miller_1995_flow}.
Applying Theorem~\ref{thm:reduction_to_circulation} proves finally the claim.
\end{proof}

By a classical characterization for feasibility of circulation problems (see
\cite{ahuja_1993_network}) the following theorem follows.

\begin{theorem}\label{thm:circulation_cut}
For any interdiction set $R$ in $G$ we have:
\begin{equation*}
\text{There exists no circulation in $\widehat{G}(R)$ $\Leftrightarrow$ there exists a
cut in $\widehat{G}(R)$ with value $<0$.}
\end{equation*}
\end{theorem}

Furthermore, as the arcs contained in a cut can be partitioned in groups of arcs
corresponding to elementary cuts, we have that there is a cut in $\widehat{G}(R)$
with value strictly less than zero exactly when there is an elementary cut in
$\widehat{G}(R)$ with value strictly less than zero. In the following, we
show how the problem of finding an interdiction set $R$ and an elementary
cut in $\widehat{G}(R)$ with negative value can be mapped onto a modified dual
network of $G$. Let
$\widetilde{G}^*=(\widetilde{V}^*,\widetilde{E}^*,\widetilde{\lambda}^*,\widetilde{c}^*)$
be the modified dual network for the network $\widehat{G}$ as introduced in
Section~\ref{sec:vertex_interdiction} \footnote{Using the notation
$\widetilde{\widehat{G}}^*$ instead of $\widetilde{G}^*$ would be more consistent.
To simplify notations we chose the second form.}. Analogously as in
Section~\ref{sec:vertex_interdiction}, we use the notations
$\widetilde{V}^*=V\cup \widehat{V}^*$ and
$\widetilde{E}^*=\widehat{E}^*\cup \widetilde{E^V}$.
We associate with every circuit $\widetilde{\mathcal{C}}\in \widetilde{G}$ a modified reduced 
length $\widetilde{\gamma}_B^*(\widetilde{\mathcal{C}})$ defined as follows. Let
$V_{\widetilde{\mathcal{C}}^*}^I \subset V$ be all the vertices of $V$ which are
surrounded in counterclockwise sense by the circuit $\widetilde{\mathcal{C}}^*$ in the graph $\widetilde{G}^*$ and let
$V_{\widetilde{\mathcal{C}}^*}\subset V_{\widetilde{\mathcal{C}}^*}^I$ be the vertices
of $V$ on the circuit $\widetilde{\mathcal{C}}^*$.
$\mathcal{T}^+_{\widetilde{\mathcal{C}}^*}$ respectively $\mathcal{T}^-_{\widetilde{\mathcal{C}}^*}$
denote the set of arcs in $\mathcal{T}$ going out respectively entering the set
$V_{\widetilde{C}^*}^I$ in $\widehat{G}$ and being adjacent to one of the vertices
in $V_{\widetilde{C}^*}$, i.e.,
\begin{align*}
\mathcal{T}^+_{\widetilde{\mathcal{C}}^*}&=\mathcal{T}\cap
\omega^+_{\widehat{G}}(V_{\widetilde{\mathcal{C}}^*}^I)\cap
\omega^+_{\widehat{G}}(V_{\widetilde{\mathcal{C}}^*})\;,\\
\mathcal{T}^-_{\widetilde{\mathcal{C}}^*}&=\mathcal{T}\cap
\omega^-_{\widehat{G}}(V_{\widetilde{\mathcal{C}}^*}^I)\cap
\omega^-_{\widehat{G}}(V_{\widetilde{\mathcal{C}}^*})\;.
\end{align*}

We finally define
\begin{equation}\label{eq:gamma}
\widetilde{\gamma}_B^*(\widetilde{\mathcal{C}}^*)=\widetilde{\lambda}^*_B(\widetilde{\mathcal{C}}^*)+
\widehat{u}(\mathcal{T}^+_{\widetilde{\mathcal{C}}^*})-
\widehat{l}(\mathcal{T}^-_{\widetilde{\mathcal{C}}^*})\;.
\end{equation}

The following theorem shows how the adapted reduced cost can be used to reformulate
the interdiction problem.

\begin{theorem}\label{thm:nfs_vi_circuit}
Let $B$ be a fixed budget. The following statements are equivalent.
\begin{enumerate}[i)]
\item There exists an interdiction set $R$ in $\widehat{G}$ such
that there exists an elementary
cut $[V',V\setminus V']$ in $\widehat{G}(R)$ with $\nu^{\widehat{G}(R)}([V',V\setminus V'])<0$.
\item There exists a circuit $\widetilde{\mathcal{C}}^*$ in $\widetilde{G}^*$ with
$\widetilde{\gamma}_B^*(\widetilde{\mathcal{C}}^*)<0$.
\end{enumerate}
\end{theorem}

\begin{proof}
Let $R$ be an interdiction set in $\widehat{G}$ and $[V',V\setminus V']$ be an elementary cut in
$\widehat{G}(R)$ with strictly negative value. We partition the set $R$ into
$R=V_R\cup \widehat{E}_R$ with $V_R=R\cap V$ and
$\widehat{E}_R=R\cap \widehat{E}$. Let 
$U_+=\mathcal{T}\cap \omega^+_{\widehat{G}}(V')\cap \omega^+_{\widehat{G}}(V_R)$ respectively
$U_-=\mathcal{T}\cap \omega^-_{\widehat{G}}(V')\cap \omega^-_{\widehat{G}}(V_R)$ the
subset of arcs in $\mathcal{T}$ going out respectively entering into the set $V'$ and being adjacent
to vertices in $V_R$. Furthermore let $W=\omega^+_{\widehat{G}\setminus R}(V'\setminus R)$. The value
of the cut $[V',V\setminus V']$ in $\widehat{G}(R)$ can be rewritten in the following way.
\begin{equation*}
\nu^{\widehat{G}(R)}([V',V\setminus V'])=\widehat{u}(W)+\widehat{u}(U_+)-\widehat{l}(U_-)
\end{equation*}

Notice that in the network $\widehat{G}\setminus (R\cup W)$, there is no path from
$V'$ to vertices in $V\setminus V'$. This implies that we can find a counterclockwise
circuit $\widetilde{\mathcal{C}}^*$ in $\widetilde{G}^*$ consisting only of dual arcs of arcs in
$\widehat{E}_R \cup W$ and of arcs being adjacent to vertices in $V_R$. Such a circuit furthermore satisfies
$V_{\widetilde{\mathcal{C}}^*}^I=V'$. This implies $U_+=\mathcal{T}^+_{\widetilde{\mathcal{C}}^*}$ and
$U_-=\mathcal{T}^-_{\widetilde{\mathcal{C}}^*}$. By a reasoning identical to the one in
Section~\ref{sec:vertex_interdiction} we have
$\widetilde{\lambda}^*_B(\widetilde{\mathcal{C}}^*)\leq \widehat{u}(W)$ implying finally 
$\widetilde{\gamma}^*_B(\widetilde{\mathcal{C}}^*)\leq \nu^{\widehat{G}(R)}([V',V\setminus V'])<0$.

Conversely let $\widetilde{\mathcal{C}}^*$ be a circuit in $\widetilde{G}^*$ with
$\widetilde{\gamma}^*_B(\widetilde{C}^*)<0$. We will show how to find an interdiction
set $R$ and a cut $[V',V\setminus V']$ in $\widehat{G}(R)$ with
$\nu^{\widehat{G}(R)}([V',V\setminus V'])<0$. The existence of such a cut implies directly
the existence of an elementary cut satisfying the claim. We set $V'=V_{\widetilde{\mathcal{C}}^*}^I$
and $V_R=V_{\widetilde{\mathcal{C}}^*}$. As before by defining 
$U_+=\mathcal{T}\cap \omega^+_{\widehat{G}}(V')\cap \omega^+_{\widehat{G}}(V_R)$ and
$U_-=\mathcal{T}\cap \omega^-_{\widehat{G}}(V')\cap \omega^-_{\widehat{G}}(V_R)$ 
we have $U_+=\mathcal{T}^+_{\widetilde{\mathcal{C}}^*}$ and
$U_-=\mathcal{T}^-_{\widetilde{\mathcal{C}}^*}$.
Let $\widehat{E}^*_{\widetilde{\mathcal{C}}^*} \subset \widehat{E}^*$ be the subset of all arcs
in $\widehat{E}^*$ which are contained in the circuit $\widetilde{\mathcal{C}}^*$, this corresponds
to the set of all arcs in $\widetilde{\mathcal{C}}^*$ that are not adjacent to a vertex in $V_R$.
Furthermore we define $\widehat{E}_{\widetilde{\mathcal{C}}^*}=\{\widehat{e}\in \widehat{E}\mid \widehat{e}^D\in %
\widehat{E}^*_{\widetilde{\mathcal{C}}^*}\}$. The minimal value of the cut $[V',V\setminus V']$ in $\widehat{G}(R)$
with respect to the interdiction set $R$ can be reformulated as follows.

\begin{equation}\label{eq:proof:to_dual}
\begin{aligned}
\min_{R\subset V\cup\widehat{E}: \widehat{c}(R)\leq B}\nu^{\widehat{G}(R)}([V',V\setminus V'])%
&\leq \min_{\widehat{E}_R\subset \widehat{E}:\widehat{c}(\widehat{E}_R)\leq B-\widehat{c}(V_R)}%
\nu^{\widehat{G}(V_R\cup \widehat{E}_R)}([V',V\setminus V'])\\
&=\nu_{B-\widehat{c}(V_R)}^{\widehat{G}}(E_{\widetilde{\mathcal{C}}^*})+\widehat{u}(U_+)-\widehat{l}(U_-)\\
&=\widetilde{\lambda}^*_{B-\widehat{c}(V_R)}(\widehat{E}^*_{\widetilde{C}^*})+\widehat{u}(U_+)-\widehat{l}(U_-)
\end{aligned}
\end{equation}

Note that we have furthermore
$\widetilde{\lambda}^*_{B-\widehat{c}(V_R)}(\widehat{E}^*_{\widetilde{C}^*})= \widetilde{\lambda}^*_B(\widetilde{\mathcal{C}})$.
This comes from the facts that all arcs $\widetilde{e}^*\in\widetilde{\mathcal{C}}^* \setminus \widetilde{E}^*_{\widetilde{\mathcal{C}}^*}$
satisfy $\widetilde{\lambda}^*(\widetilde{e}^*)=\infty$ and thus have to be interdicted in the calculation of 
$\widetilde{\lambda}^*_B(\widetilde{\mathcal{C}})$ and that
$\widetilde{c}^*(\widetilde{\mathcal{C}}^*\setminus \widetilde{E}^*_{\widetilde{\mathcal{C}}^*})=c(V_R)$ as already
seen in Section~\ref{sec:vertex_interdiction}.
Further developing Equation~\ref{eq:proof:to_dual} by using this relation, the definition of
$\widetilde{\gamma}^*_B$ and $U_+=\mathcal{T}^+_{\widetilde{\mathcal{C}}^*}$,
$U_-=\mathcal{T}^-_{\widetilde{\mathcal{C}}^*}$ we finally get
\begin{equation*}
\min_{R\subset V\cup\widehat{E}: \widehat{c}(R)\leq B}\nu^{\widehat{G}(R)}([V',V\setminus V'])%
\leq \widetilde{\gamma}^*_B(\widetilde{\mathcal{C}}^*)<0
\end{equation*}
proving the claim.
\end{proof}

In the next step, we introduce a new length function $\widetilde{\chi}^*:E^*\rightarrow \mathbb{Z}\cup \{\infty\}$,
which is a slight adaption of $\widetilde{\lambda}^*$ and satisfies
$\widetilde{\chi}^*_B(\widetilde{\mathcal{C}}^*)=\widetilde{\gamma}^*_B(\widetilde{\mathcal{C}}^*)$
for every circuit $\widetilde{\mathcal{C}}^*$ in $\widetilde{G}^*$,
where $\widetilde{\chi}^*_B$ is the reduced length with respect to $\widetilde{\chi}^*$ and the
budget $B$. Such a length function allows us to solve the network security problem on $G$ as in the
previous sections by looking for a circuit in $\widetilde{G}^*$ with negative reduced length
with respect to $\widetilde{\chi}^*$.

Let $\widetilde{\mathcal{C}}^*$ be a circuit in $\widetilde{G}^*$.
For every arc $\widetilde{e}^*\in\widehat{E}^*$ we set
$\widetilde{\chi}^*(\widetilde{e}^*)=\widetilde{\lambda}^*(\widetilde{e}^*)$.
The value
of $\widetilde{\chi}^*$ on the arcs in $\widetilde{E}^V$ will be defined such that 
for each circuit $\widetilde{\mathcal{C}}^*$ in $\widetilde{G}^*$ and each vertex $v\in V$ the following
equality will be satisfied.
\begin{equation}\label{eq:chi_v}
\widetilde{\chi}^*(\widetilde{\mathcal{C}}^* \cap \omega_{\widetilde{G}^*}(v))=%
\widetilde{\lambda}^*(\widetilde{\mathcal{C}}^* \cap \omega_{\widetilde{G}^*}(v))%
+\widehat{u}(\mathcal{T}^+_{\widetilde{\mathcal{C}}^*}\cap \omega_{\widetilde{G}^*}(v))%
-\widehat{l}(\mathcal{T}^-_{\widetilde{\mathcal{C}}^*}\cap \omega_{\widetilde{G}^*}(v))
\end{equation}

Summing the above equation over all vertices on the circuit $\widetilde{\mathcal{C}}^*$
gives $\widetilde{\chi}^*(\widetilde{\mathcal{C}}^*\cap \widetilde{E}^V)%
=\widetilde{\lambda}^*(\widetilde{\mathcal{C}}^*\cap \widetilde{E}^V)%
+\widehat{u}(\mathcal{T}^+_{\widetilde{\mathcal{C}}^*})%
-\widehat{l}(\mathcal{T}^-_{\widetilde{\mathcal{C}}^*})$,
which implies the desired property
$\widetilde{\chi}^*_B(\widetilde{\mathcal{C}}^*)=\widetilde{\gamma}^*_B(\widetilde{\mathcal{C}}^*)%
\;\forall\:\widetilde{\mathcal{C}}^*$ circuit in $\widetilde{G}^*$. In the following, we
define $\widetilde{\chi}^*$ on the arcs $\widetilde{E}^V$ in such a way that Equation~\ref{eq:chi_v}
is satisfied.

For all arcs $\widehat{e}^*\in \widetilde{E}^V$ adjacent to a vertex in $S\cup T$ we set
$\widetilde{\chi}^*(\widetilde{e}^*)=\widetilde{\lambda}^*(\widetilde{e}^*)=\infty$.
This ensures that Equation~\ref{eq:chi_v} is satisfied for all $v\in S\cup T$.
Let $v\in V\setminus (S\cup T)$, $\mathcal{T}^+(v)$ be all arcs in $\mathcal{T}$
going out of $v$, $\mathcal{T}^-_v$ be all arcs in $\mathcal{T}$ entering $v$,
$\mathcal{T}(v)=\mathcal{T}^+_v\cup\mathcal{T}^-_v$ and let $\widetilde{E}^V_v$ be the
set of all arcs in $\widetilde{E}^V$ adjacent to $v$. Furthermore,
let $\widetilde{e}^*_v$ be an arbitrary fixed arc in $\widetilde{E}^V_v$.
We denote by $H$ the (geometric) graph $(\widetilde{V}^*,\widetilde{E}^*\cup \mathcal{T})$
obtained by adding the arcs $\mathcal{T}$ to the planar graph $(\widetilde{V}^*,\widetilde{E}^*)$.
For $\widetilde{e}^*_1,\widetilde{e}^*_2\in \widetilde{E}^V_v$, we define
$\mathcal{T}_v(\widetilde{e}^*_1,\widetilde{e}^*_2)$ to be the set of all arcs in $\mathcal{T}_v$
that are traversed in $H$ when going in counterclockwise
sense from $\widetilde{e}^*_1$ around $v$ to $\widetilde{e}^*_2$.
Furthermore, let $\mathcal{T}_v^+(\widetilde{e}^*_1,\widetilde{e}^*_2)$ respectively
$\mathcal{T}_v^-(\widetilde{e}^*_1,\widetilde{e}^*_2)$ be the arcs in
$\mathcal{T}_v(\widetilde{e}^*_1,\widetilde{e}^*_2)$, that are entering respectively
going out of $v$. To simplify notations, we define
$\alpha(\widetilde{e}^*_1,\widetilde{e}^*_2)=%
\widehat{u}(\mathcal{T}_v^+(\widetilde{e}^*_1,\widetilde{e}^*_2))%
-\widehat{l}(\mathcal{T}_v^-(\widetilde{e}^*_1,\widetilde{e}^*_2))$.
Finally we define for $\widetilde{e}^*\in \widetilde{E}^V_v$
\begin{equation*}
\widetilde{\chi}^*(\widetilde{e}^*)=
\begin{cases}
\widetilde{\lambda}^*(\widetilde{e}^*)+\alpha(\widetilde{e}^*,\widetilde{e}^*_v)%
&\text{if $\widetilde{e}^*$ enters $v$}\\
\widetilde{\lambda}^*(\widetilde{e}^*)-\alpha(\widetilde{e}^*,\widetilde{e}^*_v)%
&\text{if $\widetilde{e}^*$ leaves $v$\;.}
\end{cases}
\end{equation*}

Let $\widetilde{\mathcal{C}}^*$ be a circuit in $\widetilde{G}^*$ and $v\in V$ a vertex though which
$\widetilde{\mathcal{C}}^*$ passes. We denote by $\widetilde{e}^*_{\mathrm{in}}$ respectively
$\widetilde{e}^*_{\mathrm{out}}$
the incoming and outgoing arc of $\widetilde{\mathcal{C}}^*$ with respect to the vertex $v$.
Using the introduced notation, Equation~\ref{eq:chi_v} can be rewritten as
\begin{equation}\label{eq:chi_v_revisited}
\widetilde{\chi}^*(\widetilde{e}^*_{\mathrm{in}})+\widetilde{\chi}^*(\widetilde{e}^*_{\mathrm{out}})=%
\widetilde{\lambda}^*(\widetilde{e}^*_{\mathrm{in}})+\widetilde{\lambda}^*(\widetilde{e}^*_{\mathrm{out}})%
+\alpha(\widetilde{e}^*_{\mathrm{in}},\widetilde{e}^*_{\mathrm{out}})\tag{\ref{eq:chi_v}$'$}\:.
\end{equation}

Applying furthermore the definition of $\widetilde{\chi}^*$ in Equation~\ref{eq:chi_v_revisited},
we finally have to prove the following proposition.
\begin{proposition}\label{prop2:chi_v_revisited2}
\begin{equation*}
\alpha(\widetilde{e}^*_{\mathrm{in}},\widetilde{e}^*_v)-\alpha(\widetilde{e}^*_{\mathrm{out}},\widetilde{e}^*_v)=%
\alpha(\widetilde{e}^*_{\mathrm{in}},\widetilde{e}^*_{\mathrm{out}})
\end{equation*}
\end{proposition}

\begin{proof}
We distinguish two cases describing two possible constellations of the arcs
$\widetilde{e}^*_v, \widetilde{e}^*_{\mathrm{in}}, \widetilde{e}^*_{\mathrm{out}}$ around the vertex $v$.
When going in counterclockwise sense around $v$ beginning at $\widetilde{e}^*_{\mathrm{in}}$,
these arcs are either encountered in the order
$(\widetilde{e}^*_{\mathrm{in}},\widetilde{e}^*_{\mathrm{out}},\widetilde{e}^*_v)$  or
$(\widetilde{e}^*_{\mathrm{in}},\widetilde{e}^*_v,\widetilde{e}^*_{\mathrm{out}})$.

\textbf{Case 1:} $(\widetilde{e}^*_{\mathrm{in}},\widetilde{e}^*_{\mathrm{out}},\widetilde{e}^*_v)$\\
The result follows immediately by the definition of $\alpha$ and the relations
\begin{align*}
\widehat{u}(\mathcal{T}^+_v(\widetilde{e}^*_{\mathrm{in}},\widetilde{e}^*_{\mathrm{out}}))+%
\widehat{u}(\mathcal{T}^+_v(\widetilde{e}^*_{\mathrm{out}},\widetilde{e}^*_v))&=%
\widehat{u}(\mathcal{T}^+_v(\widetilde{e}^*_{\mathrm{in}},\widetilde{e}^*_v))\\%
\widehat{l}(\mathcal{T}^-_v(\widetilde{e}^*_{\mathrm{in}},\widetilde{e}^*_{\mathrm{out}}))+%
\widehat{l}(\mathcal{T}^-_v(\widetilde{e}^*_{\mathrm{out}},\widetilde{e}^*_v))&=%
\widehat{l}(\mathcal{T}^-_v(\widetilde{e}^*_{\mathrm{in}},\widetilde{e}^*_v))\:.
\end{align*}

\textbf{Case 2:} $(\widetilde{e}^*_{\mathrm{in}},\widetilde{e}^*_v,\widetilde{e}^*_{\mathrm{out}})$\\
With the same reasoning as in the first case we get
\begin{equation*}
\alpha(\widetilde{e}^*_{\mathrm{in}},\widetilde{e}^*_{\mathrm{out}})-\alpha(\widetilde{e}^*_{\mathrm{in}},\widetilde{e}^*_v)=%
\alpha(\widetilde{e}^*_v,\widetilde{e}^*_{\mathrm{out}})\:.
\end{equation*}
The proposition thus reduces to
$\alpha(\widetilde{e}^*_v,\widetilde{e}^*_{\mathrm{out}})=-\alpha(\widetilde{e}^*_{\mathrm{out}},\widetilde{e}^*_v)$, which
follows by observing that by construction of $\mathcal{T}$ and the fact that $v\not\in S\cup T$ we have
\begin{equation*}
\widehat{u}(\mathcal{T}_v^+)-\widehat{l}(\mathcal{T}_v^-)=0\;.
\end{equation*}
\end{proof}

The network security problem on $G$ can thus be solved as in
Section~\ref{subsec:netsecalgo} 
by looking for a circuit $\widetilde{\mathcal{C}}^*$
in $\widetilde{G}^*$ with negative
reduced length with respect to $\widetilde{\chi}^*$. If and only if  
such a circuit exists, it is possible to diminish the
value of a maximum flow in the
network $G$ by at least one unit given the budget $B$.
It follows from the proof of Theorem~\ref{thm:nfs_vi_circuit}
that in this case, an interdiction strategy can be found in the
same way as in presented in Section~\ref{sec:vertex_interdiction}. 
As the network $\widetilde{G}$ used in this section is only at most
a constant factor larger than the auxiliary network used in
Section~\ref{subsec:netsecalgo}, the algorithm can be implemented
with a running time of $\mathcal{O}(B n^3)$ or, by reversing the roles of budget
and length, with a running time of $\mathcal{O}(\nu^{\max}_B(G)n^3)$. Here, too,
we expect that the running time can be improved by exploiting more deeply the structure
of the network $\widetilde{G}$ in the dynamic programming steps.

\section{Conclusions}\label{sec:conclusions}

We proposed a planarity-preserving
transformation that allows to incorporate vertex removals and vertex
capacities in pseudo-polynomial interdiction algorithms for planar graphs. Additionally, 
a pseudo-polynomial algorithm was introduced for the problem of determining
the minimum interdiction budget needed to make it
impossible to satisfy the demand of all sink nodes. The algorithm works on planar networks
with multiple sources and sinks where the sum of the supplies at
the source nodes equals the sum of the demands at the sink nodes.
However this algorithm does not seem to extend easily to general network interdiction problems with
multiple sources and sinks. Thus, it is still
not known whether network flow interdiction on planar graphs with multiple sources and sinks
is solvable in pseudo-polynomial time. We showed that the $k$-densest
subgraph problem on planar graphs can be polynomially reduced to a network flow
interdiction problem on a planar graph with multiple sources and sinks.
The algorithms presented in this paper can easily be adapted to the case when multiple
resources are needed for removing arcs and nodes, still remaining pseudo-polynomial.

The main purpose of the algorithms presented in this paper, was to show that various
interdiction problems on planar graphs can be solved in pseudo-polynomial time. We
expect however, that it should be possible to speed up the proposed algorithms
by using more elaborate techniques as for example nested dissection.
Furthermore, it would be interesting to extend the presented work to
nearly planar networks as for example networks with a bounded crossing number or genus.

\bibliographystyle{plain}
\bibliography{literature}

\begin{thebibliography}{10}

\bibitem{ahuja_1993_network}
R.~K. Ahuja, T.~L. Magnanti, and J.~B. Orlin.
\newblock {\em Network flows: theory, algorithms, and applications}.
\newblock Prentice-Hall, Inc., Upper Saddle River, NJ, USA, 1993.

\bibitem{assimakopoulos_1987_network}
N.~Assimakopoulos.
\newblock A network interdiction model for hospital infection control.
\newblock {\em Computers in biology and medicine}, 17(6):413--422, 1987.

\bibitem{burch_2002_decompositionbased}
C.~Burch, R.~Carr, S.~Krumke, M.~Marathe, C.~Phillips, and E.~Sundberg.
\newblock {\em Network interdiction and stochastic integer programming},
  volume~22, chapter~3, pages 51--68.
\newblock Springer, 2003.
\newblock A decomposition-based pseudoapproximation algorithm for network flow
  inhibition.

\bibitem{corneil_1984_clustering}
D.~G. Corneil and Y.~Perl.
\newblock Clustering and domination in perfect graphs.
\newblock {\em Discrete Applied Mathematics}, 9:27--29, 1984.

\bibitem{fakcharoenphol_2001_planar}
J.~Fakcharoenphol.
\newblock Planar graphs, negative weight edges, shortest paths, and near linear
  time.
\newblock In {\em FOCS '01: Proceedings of the 42nd IEEE symposium on
  Foundations of Computer Science}, page 232, Washington, DC, USA, 2001. IEEE
  Computer Society.

\bibitem{ford_1956_maximal}
L.~R. Ford and D.~R. Fulkerson.
\newblock Maximal flow through a network.
\newblock {\em Canadian Journal of Mathematics}, 8:399--404, 1956.

\bibitem{garey_1979_computers}
M.~R. Garey and D.~S. Johnson.
\newblock {\em Computers and Intractability; A Guide to the Theory of
  NP-Completeness}.
\newblock W. H. Freeman \& Co., New York, NY, USA, 1990.

\bibitem{ghare_1971_optimal}
P.~M. Ghare, D.~C. Montgomery, and W.~C. Turner.
\newblock Optimal interdiction policy for a flow network.
\newblock {\em Naval Research Logistics Quarterly}, 18:37--45, 1971.

\bibitem{keil_1991_complexity}
J.~M. Keil and T.~B. Brecht.
\newblock The complexity of clustering in planar graphs.
\newblock {\em J. Combinatorial Mathematics and Combinatorial Computing},
  9:155--159, 1991.

\bibitem{khuller_1994_flow}
S.~Khuller and J.~Naor.
\newblock Flow in planar graphs with vertex capacities.
\newblock {\em Algorithmica}, 11(3):200--225, 1994.

\bibitem{korte_2006_combinatorial}
B.~Korte and J.~Vygen.
\newblock {\em Combinatorial Optimization, Theory and Algorithms}.
\newblock Springer, 3 edition, 2006.

\bibitem{lawler_1976_combinatorial}
E.~L. Lawler.
\newblock {\em Combinatorial Optimization: Networks and Matroids}.
\newblock Holt, Rinehart and Winston, 1976.

\bibitem{mcmasters_1970_optimal}
A.~W. McMasters and T.~M. Mustin.
\newblock Optimal interdiction of a supply network.
\newblock {\em Naval Research Logistics Quarterly}, 17(3):261--268, 1970.

\bibitem{miller_1995_flow}
G.~L. Miller and J.~Naor.
\newblock Flow in planar graphs with multiple sources and sinks.
\newblock {\em SIAM J. Comput.}, 24(5):1002--1017, 1995.

\bibitem{phillips_1993_network}
C.~A. Phillips.
\newblock The network inhibition problem.
\newblock In {\em STOC '93: Proceedings of the twenty-fifth annual ACM
  symposium on Theory of computing}, pages 776--785, New York, NY, USA, 1993.
  ACM Press.

\bibitem{ratliff_1975_finding}
H.~D. Ratliff, G.~T. Sicilia, and S.~H. Lubore.
\newblock Finding the $n$ most vital links in flow networks.
\newblock {\em Management Science}, 21(5):531--539, 1975.

\bibitem{salmeron_2004_analysis}
J.~Salmeron, K.~Wood, and R.~Baldick.
\newblock Analysis of electric grid security under terrorist thread.
\newblock {\em IEEE Transaction on Power Systems}, 19(2):905--912, 2004.

\bibitem{wollmer_1964_removing}
R.~Wollmer.
\newblock Removing arcs from a network.
\newblock {\em Operations Research}, 12(6):934--940, 1964.

\bibitem{wood_1993_deterministic}
R.~K. Wood.
\newblock Deterministic network interdiction.
\newblock {\em Mathematical and Computer Modeling}, 17(2):1--18, 1993.

\end{thebibliography}

\end{document}